# Exotic Spin Excitation Continuum in a Weakly Coupled Quantum Chainsaw Antiferromagnet

Asiri Thennakoon[1], Prena Chaudhary[1], Sankha Subhra Bakshi[1], Tommy Park[1], Tristen Lowrey[1], Daniel Pajerowski[2], Christina Hoffmann[2], Junghong H. He[2], Hiroaki Ueda[3], Collin Broholm[4], Gia-Wei Chern[1], and Seung-Hun Lee[1]

[1]Department of Physics, University of Virginia, Charlottesville, Virginia, 22904, USA.

[2]Oak Ridge National Laboratory, Oak Ridge, Tennessee, 37831, USA

[3]Institute for Advanced Materials Research and Development, Shimane University, Shimane 690-8504, Japan.

[4]Institute for Quantum Matter and Department of Physics and Astronomy, The Johns Hopkins University; Baltimore, Maryland, 21218, USA.

## Abstract

Collective motions in strongly interacting magnets involve many spins and are often described in terms of integer-spin excitations. However, in certain cases, the collective motion can behave as if these integer excitations break apart into smaller, particle-like entities with unusual properties. Such fractionalized excitations in quantum magnets are commonly associated either with topological order in two dimensions or with criticality in one dimension. It remains unclear how these distinct mechanisms are connected across a dimensional crossover. Here we investigate the Ti-based quantum antiferromagnet, $Cs_8LiNa_3Ti_{12}F_{48}$, in which $Ti^{3+}$ ($3d^1$, $s$ = ½) ions interact antiferromagnetically within distorted kagome planes. Our inelastic neutron scattering study on a single crystal reveals a frustrated network of weakly coupled spin-1/2 chainsaws, realizing a regime of *dimensional frustration* in which interchain couplings fail to establish coherent two-dimensional order. The magnetic excitation spectrum exhibits a strong continuum spanning the full measured momentum and energy phase space. In addition, the dynamic spin correlation function displays rod-like scattering in momentum space, indicating a quasi-one-dimensional nature of the magnetic correlations. These results point to fractionalized excitations with intrinsically directional character, demonstrating that signatures of one-dimensional criticality can persist within a two-dimensional lattice. Our findings establish anisotropic fractionalization as a distinct organizing principle for quantum-disordered states.

Quantum spin liquids (QSLs) represent a striking departure from conventional phases of matter, where strong interactions and quantum fluctuations prevent symmetry-breaking order and instead produce highly entangled states with fractionalized excitations [1-5]. In two dimensions (2D), frustrated lattices such as the kagome network have been central to the search for QSLs [6-14], with much of the focus on intrinsic phases associated with emergent gauge fields and topological order [15-19]. By contrast, one-dimensional (1D) spin-1/2 systems provide a distinct and well-established route to fractionalization, where gapless spinon excitations arise in a critical quantum liquid without topological character [20-25]. A key open question is how these fundamentally different forms of quantum disorder evolve as dimensionality is tuned from one to two dimensions.

A particularly revealing regime emerges when gapless 1D spin liquids are coupled through frustrated interactions. In such systems, interchain coupling tends to promote magnetic order, while frustration suppresses coherent ordering across chains. This "frustration of dimensionality" [26-29] can prevent the system from fully developing 2D order, stabilizing instead an intermediate quantum state that is neither a set of decoupled chains nor an isotropic 2D spin liquid. The resulting state retains the criticality and fractionalization of one dimension, yet exhibits intrinsically higher-dimensional correlations, placing it outside conventional theoretical descriptions of both ordered magnets and isotropic QSLs.

This regime is expected to leave clear signatures in the excitation spectrum. In contrast to conventional kagome QSLs, where fractionalized excitations typically form broadly isotropic continua in momentum space, a state built from frustrated coupled chains should exhibit strongly anisotropic fractionalization. Spinon-like excitations remain coherent along the dominant chain direction but become incoherent or strongly renormalized across chains, leading to direction-dependent continua and pronounced redistribution of spectral weight. Observing such anisotropic fractionalization provides direct insight into the dimensional crossover from one-dimensional criticality to higher-dimensional quantum matter.

More broadly, these considerations highlight a class of quantum disordered states in which fractionalization is not tied to topological order. While topological QSLs have dominated the discussion of fractionalization in two dimensions, one-dimensional systems demonstrate that fractionalization can also arise from criticality. Extending this mechanism into two dimensions represents a fundamentally different route to quantum spin liquids, one that remains largely unexplored experimentally.

Here we report a neutron-scattering study of a new Ti-based spin-1/2 kagome compound, $Cs_8LiNa_3Ti_{12}F_{48}$, that realizes this regime. Ti-based kagome fluorides have recently emerged as a promising platform for spin-1/2 kagome magnetism, offering tunable exchange interactions and a broad landscape of magnetic ground states [14,30-34]. Although structurally derived from the kagome lattice, the magnetic interactions in $Cs_8LiNa_3Ti_{12}F_{48}$ are strongly anisotropic, effectively organizing the system into frustrated networks of coupled sawtooth-like chains embedded in two dimensions. Using time-of-flight inelastic neutron scattering on single crystals, we map out the dynamic spin correlations over a broad range of energies and momenta. The data reveal highly anisotropic excitation continua that differ qualitatively from those of a uniform kagome antiferromagnet. While anisotropic exchange models capture the leading directional features, semiclassical spin-wave descriptions fail to account for the observed spectrum. Instead, the response points to fractionalized excitations with a clear directional character, providing evidence for anisotropic fractionalization in a two-dimensional setting. These results establish $Cs_8LiNa_3Ti_{12}F_{48}$ as a platform for frustration-driven dimensional crossover and demonstrate that fractionalization in two dimensions need not rely on topological order.

The crystal structure of $Cs_8LiNa_3Ti_{12}F_{48}$ was refined utilizing room-temperature neutron diffraction measurements on a single crystal, performed on the high-resolution single-crystal neutron diffractometer TOPAZ located at the Spallation Neutron Source (SNS). The structure was refined in the monoclinic space group $Cm$, with lattice parameters $a = 14.7933(15)$ Å, $b = 15.3710(30)$ Å, $c = 10.5603(12)$ Å, $\alpha = \gamma = 90.0°$, and $\beta = 91.442(10)°$. The atomic parameters yielding the best fit, as shown in Fig. 1a, are listed in Table 1 of the Methods section.

The non-magnetic Cs, Li, and Na ions occupy the interlayer regions between Ti−F sheets, while the magnetic Ti sublattice forms distorted kagome-like planes. Each Ti ion is coordinated by six F ligands, forming distorted $TiF_6$ octahedra (See the Supplementary Information for details). The presence of multiple inequivalent $Ti^{3+}$ sites and nonuniform Ti–F coordination implies that the Ti–F–Ti superexchange pathways are not symmetry-equivalent. Accordingly, the material is not expected to realize a uniform nearest-neighbour kagome Hamiltonian. Rather, its structure naturally suggests strongly bond-dependent in-plane interactions. The nearest-neighbor Ti−Ti distance within a Ti-based layer is approximately 3.7 Å, whereas adjacent layers are separated by about 6.1 Å. This geometry is consistent with predominantly two-dimensional magnetic interactions accompanied by substantial in-plane anisotropy, as illustrated in Fig. 1b.

The bulk magnetic susceptibility $\chi(\mathrm{T})$, measured with the external magnetic field of $\mu_0 H = 1$ T applied parallel and perpendicular to the kagome plane, is shown in Fig. 2. The data for the two orientations nearly overlap over the full measured temperature range, down to approximately 2 K, indicating only weak bulk magnetic anisotropy. Upon cooling, $\chi(\mathrm{T})$ increases smoothly without any sharp anomaly, demonstrating that no conventional magnetic phase transition is detected down to 2 K. As shown in the inset, the high-temperature inverse susceptibility is approximately linear and is well described by the Curie-Weiss law. The best fit yields $\theta_{CW} = -36.91$ K (comparable to that of $Cs_8RbK_3Ti_{12}F_{48}$, $-47.4$ K) and an effective moment $p_{eff} = 1.693\ \mu_B$. The negative Curie-Weiss temperature indicates dominant antiferromagnetic interactions, while the effective moment lies close to the spin-only value $g\sqrt{S(S+1)}\mu_B = \sqrt{3}\mu_B \approx 1.73\mu_B$ expected for $g = 2$ and $S = \frac{1}{2}$, consistent with magnetism arising primarily from $Ti^{3+}$ moments. If one interprets the Curie-Weiss scale in terms of a two-dimensional nearest-neighbour kagome antiferromagnet, it corresponds to an average exchange scale of $\bar{J} \approx 3.18$ meV. Because $Cs_8LiNa_3Ti_{12}F_{48}$ is structurally distorted and magnetically anisotropic, however, this value should be regarded only as an average interaction scale rather than a microscopic Hamiltonian parameter.

Interestingly, $\chi$(T) of $Cs_8LiNa_3Ti_{12}F_{48}$ does not exhibit the pronounced broad maximum at low temperatures that was observed near 12 K in $Cs_8RbK_3Ti_{12}F_{48}$ [14], a feature that was qualitatively accounted for by theoretical models of a uniform quantum kagome antiferromagnet [35,36]. Instead, $\chi$(T) of $Cs_8LiNa_3Ti_{12}F_{48}$ shows a gradual low-temperature upturn, suggesting that the underlying magnetic correlations differ substantially from those expected for a quantum kagome antiferromagnet. This contrast calls for more sophisticated microscopic probes to elucidate the nature of the magnetic correlations in $Cs_8LiNa_3Ti_{12}F_{48}$.

For this purpose, time-of-flight neutron scattering measurements were performed on a $^7$Li-enriched single crystal of $Cs_8LiNa_3Ti_{12}F_{48}$ with a mass of approximately 1.7 g, at 1.7 K and 30 K, using incident neutron energies $E_i$ = 3.32 meV and 18 meV, on the Cold Neutron Chopper Spectrometer (CNCS) at the Spallation Neutron Source. These settings provided elastic energy resolutions of approximately 0.07 and 1.00 meV, respectively. The sample was aligned in the (*HK*0) scattering plane and rotated through 180° to obtain broad coverage of $\mathbf{Q}-\omega$ space.

To elucidate the overall spatial character of the magnetic correlations, we integrated the dynamic spin correlation function $S(\mathbf{Q},\hbar\omega)$ over the full energy-transfer window to obtain the equal-time response function, $S(\mathbf{Q}) = \int_{0.2meV}^{13meV} S(\mathbf{Q},\hbar\omega)\, d(\hbar\omega)$, shown in Fig. 3. Unlike the nearly two-dimensional response with hexagonal six-fold symmetry reported previously for $Cs_8RbK_3Ti_{12}F_{48}$ [14], $S(\mathbf{Q})$ of $Cs_8LiNa_3Ti_{12}F_{48}$ does not exhibit six-fold symmetry. Instead, it displays rod-like diffuse features extending along the $\mathrm{M}_1 \to \Gamma_1 \to \mathrm{M}_1' \to \Gamma_2$ direction in reciprocal space (see Figure 3a). This rod-like intensity indicates that the scattering varies only weakly with momentum along one direction, while remaining more strongly modulated in the perpendicular direction. Although the underlying Ti network is kagome-based, the instantaneous magnetic response is therefore far closer to a quasi-one-dimensional form than to that expected for a uniform two-dimensional kagome antiferromagnet.

Figure 4 shows the energy-resolved dynamic spin correlation function $S(\mathbf{Q},\hbar\omega)$ measured along three representative momentum cuts through the two-dimensional Brillouin zone: $\mathrm{M}_1 \to \Gamma_1 \to \mathrm{M}_1' \to \Gamma_2$(parallel to the rod-like intensity of $S(\mathbf{Q})$), $\Gamma_2 \to \Gamma_0$, and $\Gamma_0 \to \mathrm{K}_0' \to \mathrm{M}_1' \to \mathrm{K}_1 \to \Gamma_{12} \to \mathrm{K}_2'$ (perpendicular to the rod-like intensity), as indicated in the inset of Fig. 4a. The most striking feature of $S(\mathbf{Q},\hbar\omega)$ is the presence of a strong excitation continuum spanning the full measured momentum range and extending in energy up to approximately $\hbar\omega \sim 11$ meV. Another salient feature is that, at 1.7 K, $S(\mathbf{Q},\hbar\omega)$ exhibits intense low-energy excitations for $\hbar\omega \lesssim 3$ meV along the $\mathrm{M}_1 \to \Gamma_1 \to \mathrm{M}_1' \to \Gamma_2$ direction, which is parallel to the rod-like $S(\mathbf{Q})$, and near the $\mathrm{M}_1'$ point along the $\Gamma_0 \to \mathrm{K}_0' \to \mathrm{M}_1' \to \mathrm{K}_1 \to \Gamma_{12} \to \mathrm{K}_2'$ trajectory, which is perpendicular to the rod-like $S(\mathbf{Q})$. These strong low-energy excitations weaken and broaden in $\hbar\omega$ at 30 K, confirming its magnetic origin.

In other words, at 1.7 K, along the $\mathrm{M}_1 \rightarrow \Gamma_1 \rightarrow \mathrm{M}_1' \rightarrow \Gamma_2$ direction, $S(\mathbf{Q}, \hbar\omega)$ is only weakly dispersive. By contrast, along the $\Gamma_0 \rightarrow \mathrm{K}_0' \rightarrow \mathrm{M}_1' \rightarrow \mathrm{K}_1 \rightarrow \Gamma_{12} \rightarrow \mathrm{K}_2'$ direction, $S(\mathbf{Q}, \hbar\omega)$ evolves much more strongly with changing momentum, indicating pronounced dispersive behavior. As shown in Fig. 4b, this dispersion is even more clearly resolved along the $\mathrm{K}_2 \rightarrow \Gamma_2 \rightarrow \mathrm{M}_0 \rightarrow \Gamma_2 \rightarrow \mathrm{K}_2$ trajectory, which is likewise perpendicular to the rod-like $S(\mathbf{Q})$. These observations confirm the quasi-one-dimensional nature of the magnetic correlations, propagating primarily along the $\mathrm{M}_1 \rightarrow \Gamma_1 \rightarrow \mathrm{M}_1' \rightarrow \Gamma_2$ direction in momentum space of $Cs_8LiNa_3Ti_{12}F_{48}$.

To gain insight into the real-space spin correlations underlying the observed $S(\mathbf{Q}, \hbar\omega)$, we first employed linear spin wave (LSW) theory. Although LSW theory cannot account for the broad excitation continuum, it can nevertheless provide a phenomenological and minimal effective spin Hamiltonian that qualitatively reproduces the weakly dispersive behavior along the $\mathrm{M}_1 \rightarrow \Gamma_1 \rightarrow \mathrm{M}_1' \rightarrow \Gamma_2$ direction and the more strongly dispersive response along the $\mathrm{K}_2 \rightarrow \Gamma_2 \rightarrow \mathrm{M}_0 \rightarrow \Gamma_2 \rightarrow \mathrm{K}_2$ direction.

The resulting minimal spin Hamiltonian contains four distinct nearest neighbor exchange couplings, $J_\mu$ $(\mu = A, B, C, D)$ within the kagome plane, as shown in Fig. 1c. The corresponding out-of-plane Dzyaloshinskii-Moriya interactions are denoted by $D_\mu$ $(\mu = A, B, C, D)$:

$$\mathcal{H} = \sum_{\mu=A,B,C,D} \sum_{\langle ij \rangle} \left[ J_\mu \vec{S}_i \cdot \vec{S}_j + D_\mu \hat{z} \cdot \left( \vec{S}_i \times \vec{S}_j \right) \right]$$

The best-fit parameters are $J_A = 11.26$ meV, $J_B = 3.20$ meV, $J_C = 8.15$ meV, and $J_D = 1.22$ meV, together with $D_A = 0.26$ meV, $D_B = 0.17$ meV, $D_C = 0.22$ meV, and $D_D = 0.08$ meV. The black lines overlaid on the upper panels of Fig. 4a and 4b represent the corresponding calculated spin wave dispersions. Fig. 3b shows the equal-time response function $S_{LSW}(\mathbf{Q})$ obtained from the linear spin-wave calculations, which qualitatively reproduces the salient rod-like scattering along the $\mathrm{M}_1 \rightarrow \Gamma_1 \rightarrow \mathrm{M}_1' \rightarrow \Gamma_2$ directions.

The spin-wave calculated dynamic spin correlation function $S_{LSW}(\mathbf{Q}, \hbar\omega)$, on the other hand, fails entirely to reproduce the broad excitation continuum in energy at all, as shown in the lower panels of Fig. 4a and 4b. These results demonstrate that a simple semiclassical spin-wave description is wholly insufficient to account for the observed continuum, which extends throughout the full $\mathbf{Q} - \hbar\omega$ phase space.

The neutron-scattering results establish $Cs_8LiNa_3Ti_{12}F_{48}$ as a rare realization of a highly anisotropic quantum-disordered state on a kagome-derived lattice. Despite the absence of magnetic order down to 1.7 K, the spin dynamics exhibit a pronounced dichotomy: a quasi-one-dimensional structure in momentum space coexisting with an intrinsically two-dimensional lattice. The excitation spectrum is dominated by a broad continuum extending across the Brillouin zone and up to ~11 meV, which cannot be accounted for within a semiclassical description. While an anisotropic exchange model reproduces the leading directional features and bandwidth, linear spin-wave theory fails qualitatively to capture the intrinsic broadness of the response, indicating that the excitations are not conventional magnons but fractionalized degrees of freedom.

A key aspect of this system is the directional nature of the fractionalization. The rod-like equal-time correlations and weak dispersion along specific momentum directions indicate that spin correlations remain coherent along effective chain directions while becoming incoherent transverse to them. This behavior points to a form of frustration-induced dimensional reduction, in which the two-dimensional lattice dynamically reorganizes into a network of weakly coupled quantum chains. In contrast to isotropic kagome spin liquids, where fractionalization is typically momentum-independent, the present results reveal a regime of anisotropic fractionalization that lies between one-dimensional criticality and two-dimensional quantum disorder.

Several important questions remain open. It will be important to determine whether the low-energy excitations are strictly gapless or host a small but finite gap below current experimental resolution, and to clarify how continuously this state connects to one-dimensional limits such as spin-1/2 Heisenberg or sawtooth-chain models. On the theoretical side, developing a quantitative description of the momentum-resolved anisotropic continua is a key challenge. Promising directions include parton-based approaches that naturally capture fractionalization, as well as large-scale numerical methods—such as two-dimensional DMRG and tensor-network states combined with time-dependent variational principles—to directly compute dynamical structure factors in this strongly anisotropic regime.

More broadly, these results establish an alternative route to fractionalization in two dimensions that does not rely on topological order but instead emerges from the persistence of one-dimensional criticality under frustrated interchain coupling. $Cs_8LiNa_3Ti_{12}F_{48}$ thus provides a concrete experimental platform for this largely unexplored regime. More generally, the Ti-based kagome fluoride family offers a powerful setting for tuning between isotropic and quasi-one-dimensional limits, opening a pathway toward controlled studies of dimensional crossover and directionally selective fractionalization. In this sense, our results point to anisotropic fractionalization as a new organizing principle for quantum-disordered states in frustrated magnets.

# Figures

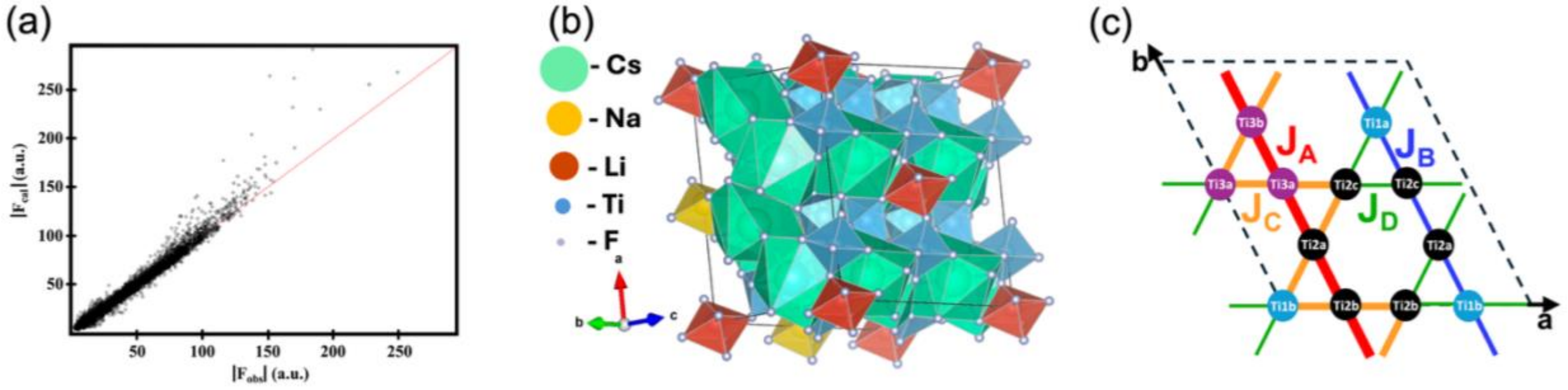

**Figure 1 | Neutron diffraction study of $Cs_8LiNa_3Ti_{12}F_{48}$.** (a) Comparison of the observed ($F_{obs}$) and calculated ($F_{calc}$) structure factors, derived from time-of-flight neutron data and crystal structure refinement using the Jana2020 package, respectively, demonstrating good agreement between experiment and the refined model. (b) Polyhedral representation of the complete three-dimensional monoclinic structure (space group $Cm$). The nonmagnetic alkali-metal cations (Cs, Na, Li) occupy the structural voids, acting as physical spacers between the magnetic Ti–F networks. (c) An isolated top-down view of the Ti sublattice highlighting a single distorted kagome layer. The fundamental repeating 12-atom titanium cluster is emphasized as dashed lines, representing one complete chemical formula unit of the kagome network.

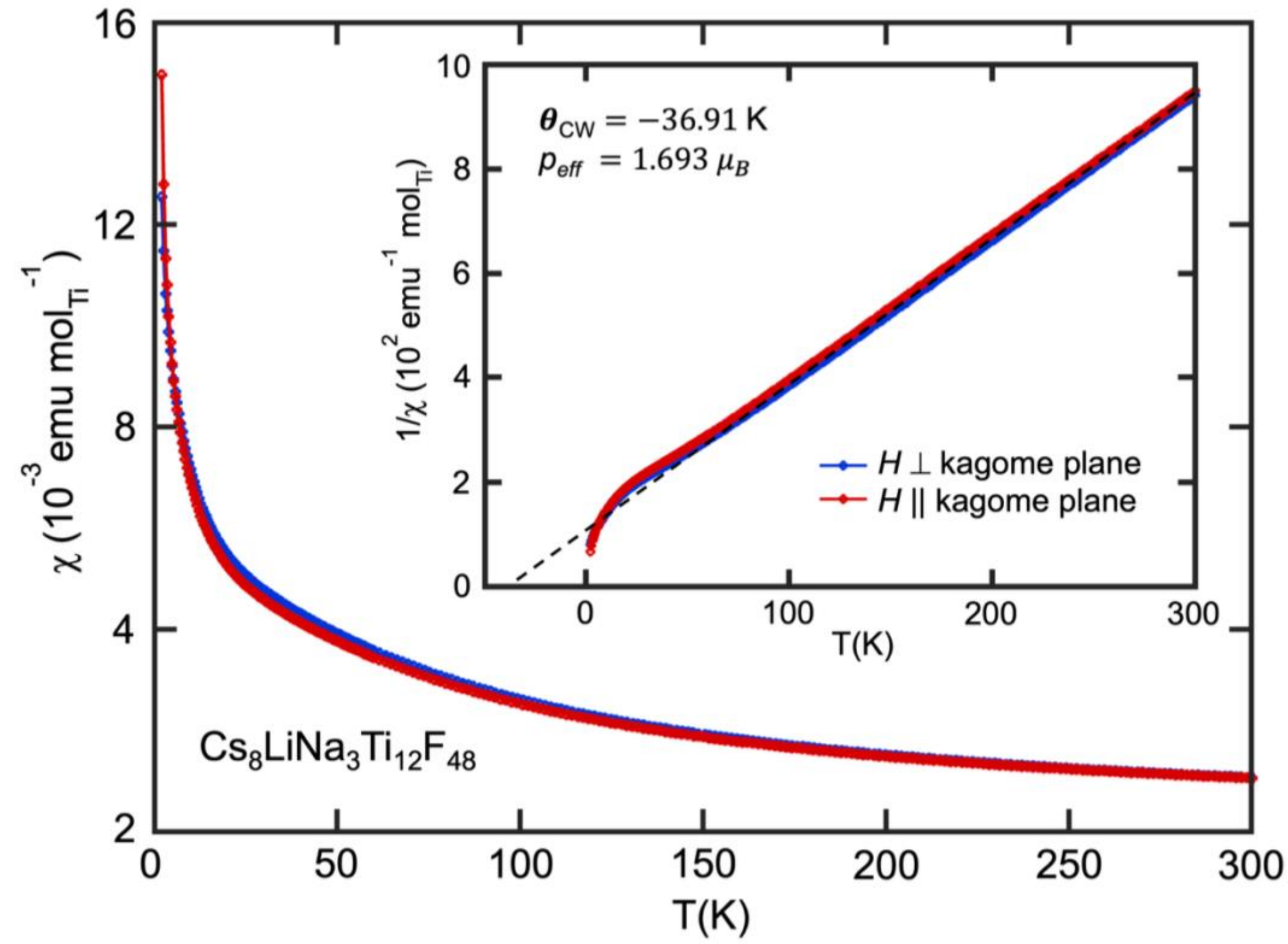

**Figure 2 | Temperature dependence of the bulk magnetic susceptibility of $Cs_8LiNa_3Ti_{12}F_{48}$,** Magnetic susceptibility, $\chi = M/H$, measured on a single crystal in an applied field of $\mu_0 H = 1$ T with the field perpendicular to the kagome plane (blue circles) and parallel to the kagome plane (red circles). The two curves nearly overlap over the full measured temperature range, indicating weak bulk magnetic anisotropy. The inset on the top right shows the inverse susceptibility $1/\chi$ together with a Curie-Weiss fit to the high-temperature region (dashed blue line), yielding a Curie-Weiss temperature $\Theta_{CW} = -36.9$ K and $p_{eff} = 1.693\ \mu_B$.

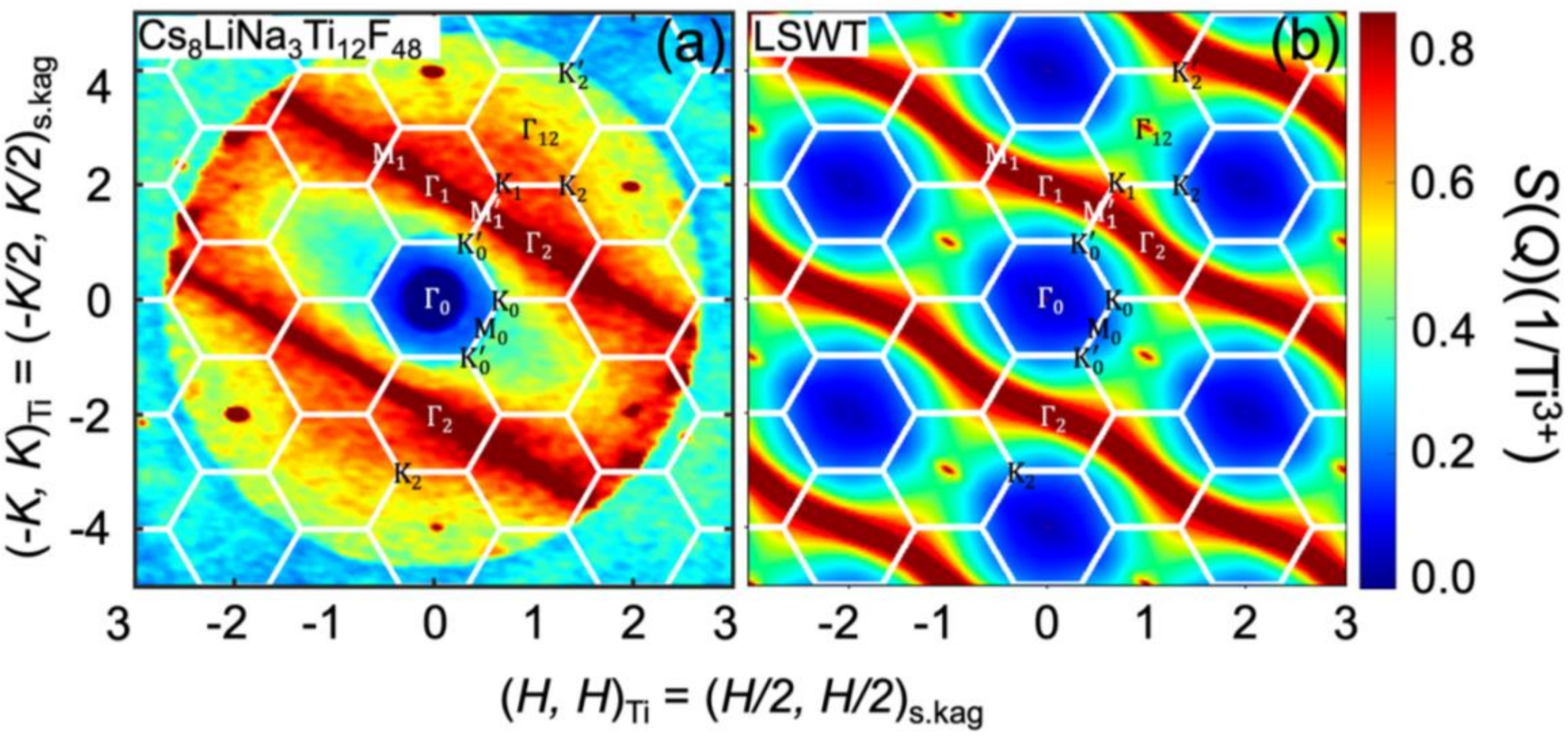

**Figure 3 | Q-dependence of the equal-time spin correlation function in $Cs_8LiNa_3Ti_{12}F_{48}$. a**, Experimental equal-time response function, $S(\mathbf{Q}) = \int_{0.2meV}^{13meV} S(\mathbf{Q}, \hbar\omega)\, d(\hbar\omega)$. Obtain at 1.7 K by integrating the inelastic neutron-scattering intensity over energy transfer. **b**, Corresponding equal-time response calculated within linear spin-wave theory (LSWT) for the anisotropic *J*-*D* model, integrated over the same energy window. The white hexagons denote repeated two-dimensional Brillouin zones, and selected high-symmetry points are labeled. The experimental map reveals strongly anisotropic rod-like diffuse scattering, whereas the LSWT result reproduces only the leading anisotropic modulation and remains substantially sharper than the data.

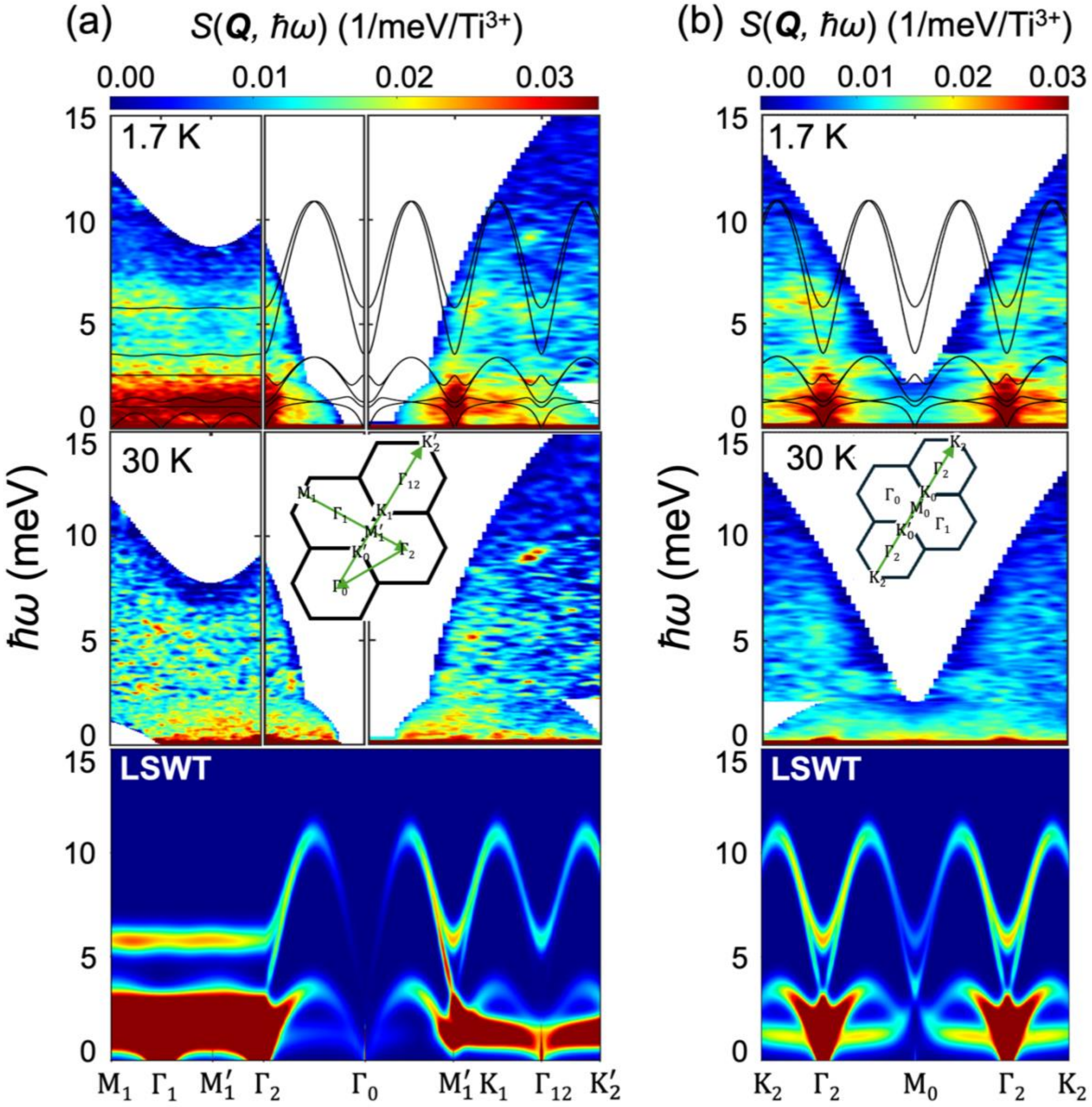

**Figure 4 | Dynamic spin correlations in $Cs_8LiNa_3Ti_{12}F_{48}$.** Color contour maps of the normalized dynamic spin correlation function, $S(\mathbf{Q}, \hbar\omega)$, as a function of momentum ($\mathbf{Q}$) and energy transfer ($\hbar\omega$), measured along two representative momentum cuts through the two-dimensional Brillouin zone, indicated by the green arrows in the insets. For each cut, the top panel shows the data at 1.7 K, the middle panel shows the data at 30 K, and the bottom panel shows the corresponding linear spin-wave theory (LSWT) intensity, $S_{LSWT}(\mathbf{Q}, \hbar\omega)$, calculated for the anisotropic *J-D* model. In the first panels, the black lines represent the calculated LSWT dispersion relation. Panel **a** follows the $\mathrm{M_1} \rightarrow \Gamma_1 \rightarrow \mathrm{M}'_1 \rightarrow \Gamma_2 \rightarrow \Gamma_0 \rightarrow \mathrm{M}'_1 \rightarrow \mathrm{K_1} \rightarrow \Gamma_{12} \rightarrow \mathrm{K}'_2$ path, while panel **(b)** shows a second cut through the $\mathrm{K_2} \rightarrow \Gamma_2 \rightarrow \mathrm{M_0} \rightarrow \Gamma_2 \rightarrow \mathrm{K_2}$ path, as indicated in the insets. The experimental spectra

are broad and strongly anisotropic, with substantial intensity distributed between the semiclassical spin-wave branches.

## Acknowledgments

A.T., P.C., S.S.B., T.P., T.L., G.W.D., and S.H.L. were supported by the U.S. Department of Energy, Office of Science, Basic Energy Sciences, under Award No. DE-SC0026087 for the project "Quantum spin states of new Ti-based kagome antiferromagnets." H.U. was supported by JSPS KAKENHI Grant Number 26K00656. This research used resources at the Spallation Neutron Source, a DOE Office of Science User Facility operated by the Oak Ridge National Laboratory. The beam time was allocated to CNCS on proposal numbers IPTS-35028.1 and IPTS-36874.1.

# Supplementary Information

## Single-Crystal Neutron Diffraction and Crystal Structure Refinement

Single-crystal neutron diffraction data for the $Cs_8LiNa_3Ti_{12}F_{48}$ single crystal were collected at 300 K using the TOPAZ time-of-flight (TOF) diffractometer at the Spallation Neutron Source. The crystal was mounted to an aluminum pin and positioned on the instrument's ambient goniometer. TOPAZ employs a broad neutron wavelength band (0.4–3.5 Å) which, together with an array of 25 detectors covering 3.2 sr in solid angle, enables highly efficient simultaneous three-dimensional reciprocal-space mapping. Data were collected over multiple goniometer settings using a series of $\omega$ rotations of the sample rotation axis about the vertical axis, spanning approximately 0° to 180° with the polar angle $\phi$ fixed at 0°, providing high redundancy and comprehensive volumetric sampling of the Bragg reflections. This extensive reciprocal space coverage was essential for accurately determining the positions of the light ions, particularly $F^-$, as the $Ti^{3+}$ ions are surrounded by six $F^-$ ligands that stabilize the kagome-related lattice. Initial data reduction, including Lorentz and absorption corrections, was carried out using the MANTID software to obtain the integrated intensities required for structural refinement.

Subsequently, structural refinement of the neutron diffraction data was performed against squared structure factor intensities ($F^2$) using the Jana2020 crystal refinement software. To properly account for wavelength-dependent corrections inherent to the TOF Laue technique, the integrated intensities were processed in the HKLF 2 format. The structural model was refined in the polar monoclinic space group $Cm$. Because this space group lacks a fixed origin along its polar axes, the lattice origin must be constrained to prevent continuous parametric drift during least-squares minimization. To anchor the refinement, the positional coordinates of the weakly scattering lithium atom were fixed at (0, 0, 0). Furthermore, because the lithium atom resides within a relatively large structural void, its thermal parameters were highly correlated with the local background. To ensure a stable and robust convergence, the isotropic displacement parameter ($U_{iso}$) of this specific lithium site was fixed to a physically reasonable value, while all other heavier atoms, including sodium, were freely refined with full anisotropic displacement parameters. The optimal parameters for the refined crystal structure are summarized in Table I and II below.

**Table I. Space group and room-temperature lattice parameters for $Cs_8LiNa_3Ti_{12}F_{48}$.** Included are the unit cell dimensions *a, b,* and *c*, the lattice angles *α, β, γ*, the unit-cell volume *V*, and the number of formula units per unit cell *Z*. Refinement statistics are also provided, including the maximum shift/error ratio, together with the residual indices defined as $R = \frac{\sum||F_{obs}| - |F_{cal}||}{\sum|F_{obs}|}$, $wR = \sqrt{\frac{\sum w(F_{obs}^2 - F_{cal}^2)^2}{\sum w(F_{obs}^2)^2}}$, and the goodness-of-fit $GoF = \sqrt{\frac{\sum w(F_{obs}^2 - F_{cal}^2)^2}{N_{obs}-N_{param}}}$ where $N_{obs}$ and $N_{param}$ denote the numbers of observed reflections and refined parameters, respectively. Numbers given in parentheses represent the standard deviations in the last significant digits.

| | $Cs_8LiNa_3Ti_{12}F_{48}$ |
|---|---|
| Space group | *Cm* |
| *a* (Å) | 14.7933(15) |
| *b* (Å) | 15.371(3) |
| *c* (Å) | 10.5603(12) |
| *α* (°) | 90.0(0) |
| *β* (°) | 91.442(10) |
| *γ* (°) | 90.0(0) |
| *V* (Å$^3$) | 2400.5(5) |
| *Z* | 2 |
| *GoF* | 1.12 |
| *R* | 6.77 |
| *wR* | 20.93 |
| Max shift/error | 0.043 |

**Table II. Atomic coordinates and equivalent isotropic displacement parameters**. The alphabetical suffixes (a, b, c) assigned to the atoms indicate how the parent atomic sites split under the monoclinic distortion. For a two-site split, **a** denotes the atom located on the mirror plane, while **b** denotes the atom occupying a general position. For a three-site split, all atoms occupy general positions and are labeled **a**, **b**, and **c**. Unsplit atoms that remain on the mirror plane are labeled **a**.

| Atom | x | y | z | Ueq (Å²) |
|---|---|---|---|---|
| Cs1a | 0.1496(3) | 0 | 0.6075(4) | 0.0270(14) |
| Cs1b | 0.39188(19) | 0.2496(3) | 0.1264(3) | 0.0276(8) |
| Cs2a | 0.6491(3) | 0 | 0.6080(4) | 0.0249(13) |
| Cs3a | 0.8972(4) | 0 | 0.3612(5) | 0.0333(17) |
| Cs3b | 0.63786(18) | 0.7501(3) | 0.8795(3) | 0.0240(7) |
| Cs4a | 0.3958(3) | 0 | 0.3612(5) | 0.0294(15) |
| Li1a | 0 | 0 | 0 | 0.0227 |
| Na1a | 0.4991(4) | 0 | 0.9930(6) | 0.0144(12) |
| Na1b | 0.28589(16) | 0.2489(3) | 0.4846(3) | 0.0115(6) |
| Ti1a | 0.2455(3) | 0 | 0.9662(5) | 0.0093(13) |
| Ti1b | 0.1430(3) | 0.8727(4) | 0.2401(4) | 0.0097(8) |
| Ti2a | 0.53857(18) | 0.2501(4) | 0.5153(2) | 0.0101(6) |
| Ti2b | 0.6430(3) | 0.8753(4) | 0.2390(4) | 0.0095(8) |
| Ti2c | 0.3941(3) | 0.8761(4) | 0.7397(4) | 0.0098(8) |
| Ti3a | 0.7443(3) | 0 | 0.9677(5) | 0.0111(14) |
| Ti3b | 0.3936(3) | 0.6234(4) | 0.7389(4) | 0.0101(8) |
| F1a | 0.7074(2) | 0.9006(3) | 0.0787(3) | 0.0261(9) |
| F1b | 0.35111(16) | 0.7502(3) | 0.7492(3) | 0.0260(7) |
| F1c | 0.50201(17) | 0.3434(2) | 0.6400(3) | 0.0199(8) |
| F2a | 0.1580(3) | 0 | 0.3010(4) | 0.0209(11) |
| F2b | 0.2056(2) | 0.1000(2) | 0.0776(3) | 0.0261(9) |
| F3a | 0.5830(2) | 0.3494(3) | 0.4048(3) | 0.0258(9) |
| F3b | 0.62693(15) | 0.7499(3) | 0.17900(18) | 0.0182(5) |
| F3c | 0.28362(18) | 0.9075(3) | 0.8417(3) | 0.0205(8) |
| F4a | 0.9354(3) | 0 | 0.7359(5) | 0.0255(12) |
| F4b | 0.78377(19) | 0.9067(2) | 0.8430(3) | 0.0221(8) |
| F5a | 0.8551(3) | 0 | 0.0558(4) | 0.0284(16) |
| F5b | 0.4619(2) | 0.6384(2) | 0.8931(3) | 0.0223(8) |
| F6a | 0.4361(3) | 0 | 0.7362(5) | 0.0256(12) |
| F6b | 0.5815(2) | 0.1508(2) | 0.4027(3) | 0.0252(9) |
| F7a | 0.65451(14) | 0.2498(3) | 0.5927(2) | 0.0318(8) |
| F7b | 0.75555(19) | 0.8479(3) | 0.3175(3) | 0.0257(9) |
| F7c | 0.4614(2) | 0.8600(3) | 0.8919(3) | 0.0240(9) |
| F8a | 0.3537(3) | 0 | 0.0570(4) | 0.0280(15) |
| F8b | 0.25513(18) | 0.8479(3) | 0.3156(3) | 0.0246(9) |
| F9a | 0.6561(3) | 0 | 0.3003(3) | 0.0211(11) |
| F9b | 0.50201(17) | 0.8421(2) | 0.6387(3) | 0.0204(8) |
| F10a | 0.52961(19) | 0.0983(2) | 0.1643(3) | 0.0237(8) |
| F10b | 0.42685(14) | 0.7496(3) | 0.4280(2) | 0.0275(7) |
| F10c | 0.3235(2) | 0.1120(2) | 0.5893(2) | 0.0245(9) |
| F11a | 0.6300(3) | 0 | 0.8874(4) | 0.0276(15) |
| F11b | 0.8239(2) | 0.1107(2) | 0.5897(2) | 0.0251(9) |
| F12a | 0.1276(3) | 0 | 0.8888(4) | 0.0278(14) |
| F12b | 0.52921(19) | 0.5973(2) | 0.1633(3) | 0.0244(9) |

**Neutron scattering measurements**

Time-of-flight inelastic neutron scattering (INS) measurements were carried out on the Cold Neutron Chopper Spectrometer (CNCS) at Oak Ridge National Laboratory (ORNL), Oak Ridge, Tennessee, USA. A single crystal with a total mass of $\sim 1.7\ g$ was mounted on Al plates using CYTOP (CTL-107M), and the plates were attached to an Al sample holder. To reduce neutron absorption, the crystal was synthesized using the $^7$Li isotope, thereby avoiding the strong absorption of natural Li arising from $^6$Li. The sample assembly was mounted in a 100 mm CNCS Orange cryostat with a base temperature of 1.7 K. The crystal was aligned in the $(HK0)$ scattering plane, with $c^*$ the crystal axis parallel to the cryostat axis and rotated through $180^0$ to achieve broad coverage of $\mathbf{Q} - \omega$ space. Measurements were performed with incident neutron energies of $E_i = 3.32$ and 18 meV, which provided an elastic line energy resolution of ~ 0.07 and ~1.00 meV, respectively. These two settings allowed us to resolve low-energy excitations while also covering the full magnetic bandwidth up to about 15 meV. Data were collected at 1.7 K and 30 K. After the sample measurements, the empty cryostat was measured under identical conditions for background subtraction. The data were reduced and analyzed using the Shiver visualization package and the Mantid Workbench software package.

**Absolute normalization of magnetic neutron scattering data**

The neutron scattering intensities were normalized to absolute units of scattering cross section by comparison with established standards, including incoherent elastic scattering from vanadium, sample incoherent elastic scattering, elastic nuclear Bragg peaks, and sample phonon scattering [37] In this work, the absolute normalization was performed using the sample incoherent elastic scattering, as described in the Supplementary Information.